**Direct imaging of dopant and impurity distributions in 2D MoS$_2$**

*Se-Ho Kim[+], Joohyun Lim[+,*], Rajib Sahu, Olga Kasian, Leigh T. Stephenson, Christina Scheu, Baptiste Gault[*]*

Se-Ho Kim, Dr. Joohyun Lim, Dr. Rajib Sahu, Dr. Olga Kasian, Dr. Leigh T. Stephenson, Prof. Christina Scheu, Dr. Baptiste Gault
Max-Planck-Institut für Eisenforschung GmbH, Max-Planck-Straße 1, Düsseldorf, 40237, Germany
E-mail: j.lim@mpie.de, b.gault@mpie.de

Dr. Baptiste Gault
Department of Materials, Royal School of Mines, Imperial College, Prince Consort Road, London, SW7 2BP, United Kingdom

Keywords: 2D nanomaterials, molybdenum disulfide, atom probe tomography

**Abstract**

Molybdenum disulfide (MoS$_2$) nanosheet is a two-dimensional material with high electron mobility and with high potential for applications in catalysis and electronics. We synthesized MoS$_2$ nanosheets using a one-pot wet-chemical synthesis route with and without Re-doping. Atom probe tomography revealed that 3.8 at.% Re is homogeneously distributed within the Re-doped sheets. Other impurities are found also integrated within the material: light elements including C, N, O, and Na, locally enriched up to 0.1 at.%, as well as heavy elements such as V and W. Analysis of the non-doped sample reveals that the W and V likely originate from the Mo precursor. We show how wet-chemical synthesis results in an uncontrolled integration of species from the solution that can affect the material's activity. Our results are expected to contribute to an improved understanding of the relationships linking composition to properties of 2D transition-metal dichalcogenide materials.

Molybdenum disulfide ($MoS_2$) is a semiconducting material; however, unlike bulk $MoS_2$ which has an indirect bandgap of 1.2 eV[1], $MoS_2$ prepared as a two-dimensional material has a direct bandgap of 1.8 eV[2] paired with a high electron mobility.[3] 2D $MoS_2$ shows advantageous properties for a wide array of applications in photo(electric)catalysis[4,5], photoluminescence[6], chemical sensors[7], electronics for field-effect transistors[8], and light harvesting devices.[9] Additionally, the oscillating piezoelectric voltage and current output[10] can be tuned by adjusting the number of stacked layers of $MoS_2$, indicating its potential application in powering nano-devices[11] and stretchable electronics.[12]

Control of the charge carrier concentration is a key to tune a material's electronic properties and its activity, and this is usually achieved by doping with the appropriate elements. For example, rhenium doping of $MoS_2$ (Re-$MoS_2$) nanosheets leads to the formation of a metallic 1T phase locally which is more conductive and catalytic active via 2D basal plane compared to the semiconducting 2H phase with edge limited catalytic activity.[13,14] However, the prospect of doped $MoS_2$ for high-performance electronics is still controversially debated.[15,16] One of the problems that hinders relating performance and doping stems from the challenge to measure the low-level concentration of p- or n-type dopants within 2D materials, which is necessary to understand their influence on the bandgap.[17] For example, Re is reported to be a n-type dopant but the local detection and quantification of Re within $MoS_2$ multi-layer is difficult.[18,19] The presence of traces of spurious elements integrated within the structure during the synthesis could also detrimentally affect their properties. Identification and quantification of these impurities are however extremely challenging. Over the past few years, both computational and experimental approaches were used to study the influence of controlled levels of various dopants to adjust the bandgap for instance in $MoS_2$ nanostructures.[20,21]

Detailed chemical and compositional information at the near atomic-scale is required for explicit understanding of concentration-property relationships of both dopant and impurity.

This is commonly gained through (scanning) transmission electron microscopy-energy-dispersive X-ray spectroscopy ((S)TEM-EDS) and X-ray photoelectron spectroscopy (XPS) to detect dopants and impurities.[22,23] Atom probe tomography (APT) represents an attractive alternative. APT is a burgeoning characterization technique allowing for mapping the elemental distribution in nanostructured materials with a unique combination of 3D capability, sub-nanometer spatial resolution, and ~10 ppm-level detection sensitivity for all elements irrespective of their mass.[24,25] APT shows great potential for nanomaterial characterization especially for dopant/impurity analysis.

Chemical vapor deposition (CVD) has been used to deposit single 2D layer or only a few stacked layers, which can be used in electronic application.[26] Less well-controlled assemblies of 2D layers synthesized by wet-chemistry have been used for catalytic applications.[27–29] Here, we studied wet-chemically self-assembled porous $MoS_2$. This material exhibits a more complex morphology compared to a simple 2D layer and could in principle be used for both electronic and catalytic applications.

First, we studied the intentional doping of $MoS_2$ with Re by using APT. We also reveal unintentional doping by impurity elements present during the synthesis and originating from the metal-precursor or the solvent. Both heavy elements, such as V and W, and light elements such as C, N, and O were reported to act as electron acceptor (p-type dopant) and Na was reported to be an n-type dopant, up to a few atomic percent of all these elements are detected in the nanosheets. Underpinned by the challenge associated with measuring low quantities of these impurity elements, the influence of the presence of both dopant and impurity elements on the activity of the material have been too often neglected. This tends to restrict strategies to optimize the materials performance to being empirical rather than guided by physics or chemistry.

We first synthesized Re-doped $MoS_2$ nanosheets by using the one-pot wet-chemical method outlined by Xia et al.[19] Ammonium molybdate (($NH_4$)$_6Mo_7O_{24}$) and thiourea ($NH_2CSNH_2$) precursors are dissolved along with ammonium perrhenate (NH$_4$ReO$_4$ for Re-doped MoS$_2$) in distilled water and heated at 200 °C for 20h. The collected powder was then thoroughly rinsed with distilled water to remove excess impurities. The details are described in the Supporting Information.

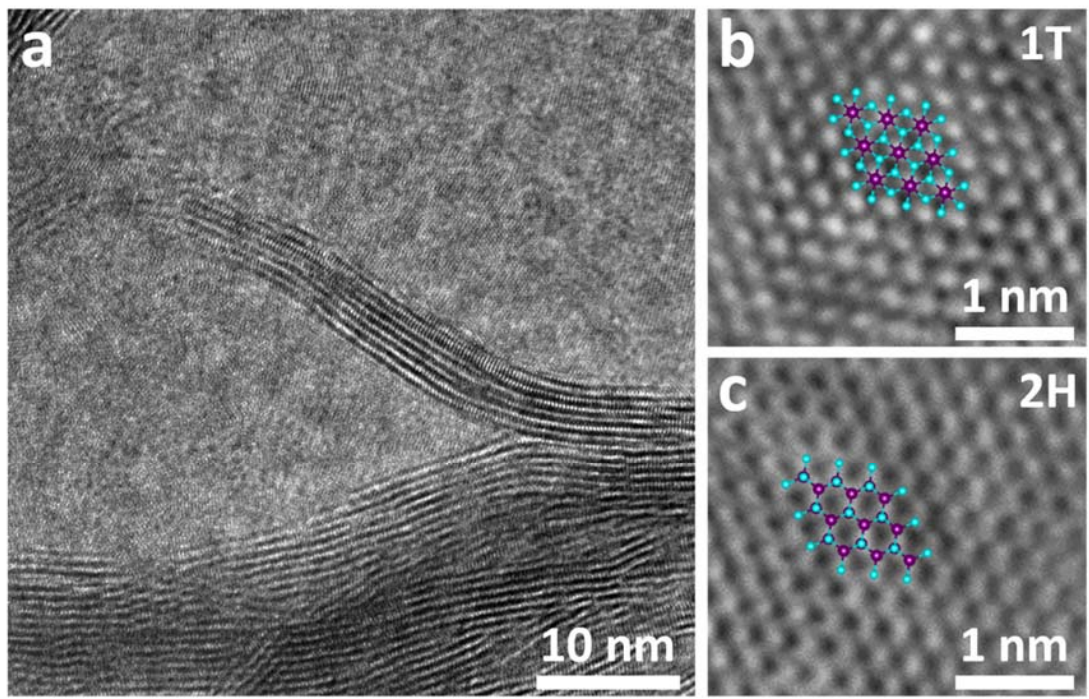

**Figure 1.** TEM images of Re-MoS$_2$ showing (a) assembled morphology in a low magnification, and local phases of (b) 1T and (c) 2H in a high magnification (purple and cyan dots represent Mo and S atoms).

Figure 1 shows transmission electron microscopy (TEM) image of the Re-doped MoS$_2$ nanosheets. We observe assemblies of 2D nanosheets, with ~6.3 nm of stacked-layers of Re-doped MoS$_2$ as displayed in Figure 1a. (S)TEM-EDS (see Figure S1a) shows that Mo, S, and Re are apparently rather homogeneously distributed within the 2D assembled nanosheets. The details on peaks in EDS spectrum is described in Figure S1b. The introduced Re induces locally the formation of the 1T phase in 2H phase of MoS$_2$ as shown in top view of high resolution TEM images (Figure 1b and c). The transition from 2H to 1T for the Re-doped MoS$_2$ is explained by crystal field theory.[14,18,19] The Fermi level of Re-doped MoS$_2$ is heavily shifted towards the conduction band minimum, and electrons are delocalized, donating electrons. These free electrons fill stable d-orbitals in the 1T-phase rather than those of 2H-phase according to the Hund's rule; therefore, the structure becomes more stable with Re content.[19]

In order to map the distribution of Re in 3D and to verify the presence of impurity elements in the Re-MoS$_2$ sheets, we performed APT. We used co-electrodeposition of a Ni-film to embed freestanding Re-MoS$_2$ nanosheets, applying the method described in Ref.[30], so as to enable the preparation of a sharp specimens (<100 nm at the apex). A similar approach had previously been used for nanoparticles[31–41], but never for 2D materials.

Scanning electron micrographs obtained from the surface of the Ni/Re-MoS$_2$ co-electroplated sample are displayed in Figure S2a and S2b. Following cross-sectional focused-ion beam milling (FIB), regions with a dark contrast appear in the micrographs in Figure S2c and S2d. These regions are ascribed to assemblies of Re-MoS$_2$ nanosheets encapsulated in the Ni-matrix, and, importantly, no noticeable voids can be detected. Site-specific APT specimen preparation was performed near one of these regions using the protocol outlined in Ref.[42] Experimental results from the Re-MoS$_2$ APT measurement are presented in Figure S3.

Figure 2a shows a reconstructed 3D atom map of the Re-MoS$_2$ nanosheets (Mo: purple, S: cyan) embedded in the Ni matrix (yellow). An iso-composition surface with a threshold of 15 at.% for Mo highlights the interface between the MoS$_2$ nanosheets and the Ni matrix within a 10 nm thin slice viewed along z-axis shows, as displayed in Figure 2b. The Re dopant atoms (in blue) clearly are partition into the nanosheets. As expected, Re-related peaks were not found in the Ni/MoS$_2$ by APT (see Figure S4) indicating that Re only originates from the Re precursor.

A cuboidal region-of-interest of the Re-MoS$_2$ region viewed along z-axis from the acquired 3D atom map (Figure 2c) reveals that Na atoms are distributed along the MoS$_2$ nanosheets. Na is not detected in the electroplated Ni (see Figure S5). Therefore, Na atoms are stemmed from the Mo precursor reagent (($NH_4$)$_6$Mo$_7$O$_{24}$, Sigma Aldrich) as it contains 0.01 wt.% of Na by the chemical specification sheet. Across the nanosheets, approx. 0.1 at.% of Na is detected and segregated within Re-MoS$_2$, forming clusters of approx. 4–5 nm in size (see Supporting Information Table S1). From the density functional theroy (DFT) calculations, Na doping was reported to enhance charge transfer of the MoS$_2$ by donating electrons causing, i.e. n-type doping.[43,44] Na is shown here incorporated within the nanosheets, and this is also rationalized by the excellent intercalation ability of MoS$_2$ for alkali ions with a high capacity and stability reported for metal-ion batteries.[45] Furthermore, it has been reported that Na atoms are detected in exfoliated-geological and CVD-grown MoS$_2$ nanosheets[46] as well as CVD-grown WS$_2$ nanosheets.[47]

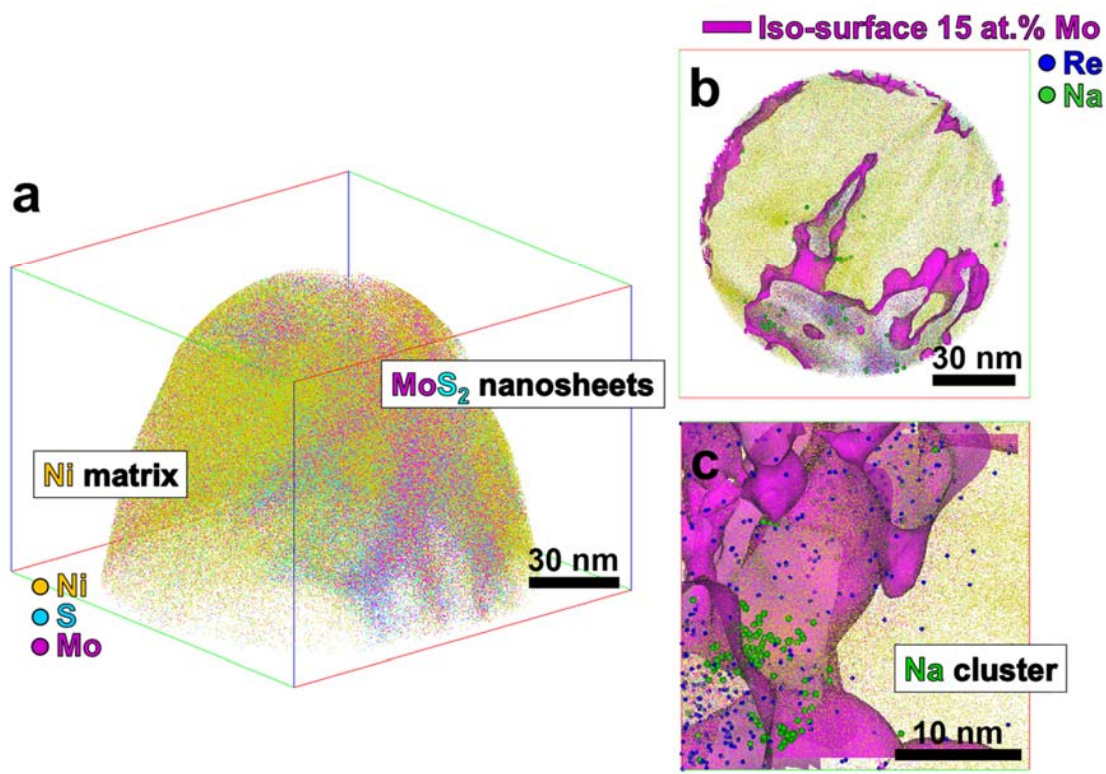

**Figure 2.** (a) 3D atom map of Re-MoS$_2$ nanosheets embedded in Ni. (b) 10 nm thin slice viewed along z-axis and (c) cubic region of interest (25 x 25 x 25 nm$^3$) sectioned from Figure 2a. The purple iso-composition surface with a thresh-hold value of 15 at.% Mo delineates the assembled MoS$_2$ nanosheets. Yellow, cyan, purple, blue, and green dots mark the reconstructed positions of Ni, S, Mo, Re, and Na atoms, respectively.

Besides undesired impurity of Na, unexpected heavy elements of V and W are also detected in both Re-doped and non-doped MoS$_2$ nanosheets. Detailed mass spectra ranges are presented in

Figure S6. Strong peaks are detected for $^{51}V^+$ and all the natural isotopes of $W^+$ (see Table S2). These elements are not found in the APT mass spectrum of the Ni matrix, i.e. where no Re-MoS$_2$ are detected (see Figure S5), whereas they are also observed in the non-doped MoS$_2$ (see Figure S7). V and W hence most likely come from the Mo precursor. According to the chemical specification sheet of the Mo precursor reagent ($(NH_4)_6Mo_7O_{24}$, Sigma Aldrich), it contains 0.001 wt.% of "heavy metals" impurities. Despite significant attention to extract high-purity Mo, the separation of Mo from V and W is reported to be very difficult, since they have similar chemical properties[48–50] and are known impurities within Mo.[51] Therefore, V and W likely remain alongside with Mo.

We performed X-ray photoelectron spectroscopy (XPS) on the undoped and Re-doped MoS$_2$ samples to look for trace amounts of V, W, and Na (see Figure S10 and S11). No other impurities except C, N, and O were found due to the detection limit of XPS. Addou et al. used inductively coupled plasma mass spectrometry (ICP-MS) and time-of-flight mass spectrometry (TOF-MS) on CVD-grown MoS$_2$ nanosheets and found V and W within MoS$_2$.[46] Likewise, they did not observe the corresponding peaks of these elements by XPS.

DFT calculation for $V_{0.08}Mo_{0.92}S_2$ showed increased electronic properties, such as 40 times increase in the in-plane conductivity and 20 times higher carrier concentration than MoS$_2$. This change in properties led to higher catalytic activity for the hydrogen evolution reaction.[54] However, the bulk atomic compositions of V and W in the Re-MoS$_2$ are only 419 and 206 ppm, respectively. The position of W is assigned to the same hexagonal parent structure of MoS$_2$ sharing the metal sites with Mo[52,53] and V doping is also considered as intralayer doping in MoS$_2$.[53] The bandgap of the semiconducting phase (2H structure) of VS$_2$ (1.87 eV)[55] and WS$_2$ (1.91 eV)[56] are similar to that of MoS$_2$ (1.78 eV)[56]. Moreover, VS$_2$, WS$_2$, and MoS$_2$ lattice constants are 0.317, 0.318, and 0.318 nm, respectively, that are very close to each other.[57] The effect of the presence of V and W in such low compositions here could have only a limited effect on the electronic properties of MoS$_2$, yet this would need to be confirmed by complementary local electronic- and catalytic property measurement.

Figure 3a shows an isolated stacked-layer of Re-MoS$_2$ nanosheets extracted from a larger reconstructed dataset containing nanosheets, evidenced by a set of iso-composition surfaces encompassing regions containing over 15 at.% Mo. The composition of the Re-doped nanosheets is extracted from a one-dimensional profile calculated within a cylindrical region-of-interest positioned perpendicular to the nanosheets, as plotted in Figure 3b. The Mo composition at the core of the MoS$_2$ nanosheets value is approx. 30 at.%. The thickness of the layer is approximately 7 nm, which agreed with TEM observations of the Re-MoS$_2$ nanosheet layers. Re atoms, shown in blue, are homogeneously distributed throughout the MoS$_2$ nanosheets. This is supported by a nearest-neighbor analysis reported in Figure S8. The composition of Re reaches 5.8 at.% in the center of the sheet and is on average 3.8 at.% within the nanosheets, suggesting that one Re atom substitutes to every 7.9 Mo atoms.

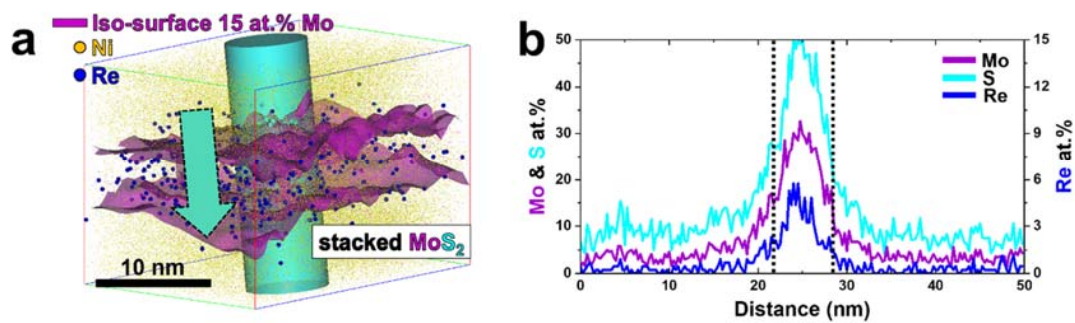

**Figure 3.** (a) 3D atom map of one stacked Re-MoS$_2$ nanosheets with iso-concentration value of Mo at 15 at.% (purple). Yellow and blue dots represent the reconstructed Ni and Re atoms, respectively. (b) 1D compositional profile across Re-MoS$_2$ nanosheets.

The atomic concentration ratio of S to Mo in the Re-doped MoS$_2$ nanosheets are 1.7 which is lower than the expected stoichiometry of MoS$_2$. It has been reported that single-layer MoS$_2$ oxidizes under ambient condition[58] as S vacancies are formed through oxidation spontaneously followed by O substitution process.[59] Using DFT calculation, the enthalpies of each reaction step were calculated to be -0.49 eV for S vacancy formation and -0.39 eV for O saturation, which implies that the oxidation is hence thermodynamically favorable.[60] Moreover, it is well known that surface of MoS$_2$ naturally oxidizes on surface.[61,62] This reaction results in a complex molecular structure of MoS$_{(2-x)}$O$_x$.[63] Peaks pertaining to MoSO$^+$ molecular ions are detected (see Figure S3b) and the 1D profile across the reconstructed MoS$_2$ nanosheets clearly show that the presence of O is not limited to the surface but rather through the MoS$_2$ assemblies as shown in Figure 4a. As a result, the stoichiometric ratio of S+O to Mo gives 1.9. Some S vacancies are expected since S vacancies are reported to be the most abundant defects in MoS$_2$[64] and Re doping, which pushes the MoS$_2$ Fermi level close to its conduction band, could result in favorable formation of S vacancies.[65] Furthermore, it is also possible that some of the S is lost during field evaporation, as has been observed for some covalently bonded materials[66], but the correlation histograms calculated here do not give indications of such specific losses as shown in Figure S9.

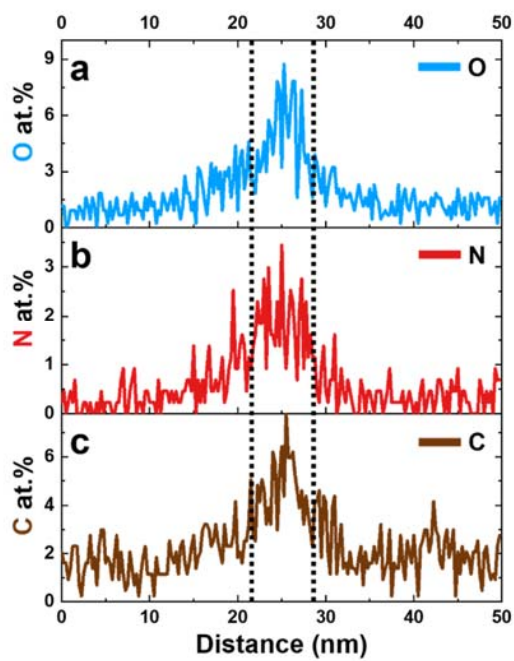

**Figure 4.** 1D atomic concentration profiles of (a) O, (b) N, and (c) C across the reconstructed Re-MoS$_2$ in Figure 3a.

The enrichment in N and C in the Re-doped $MoS_2$ nanosheets are also evidenced in Figure 4b and 4c. N and C can either originate from the precursors of Mo (($NH_4)_6Mo_7O_{24}$), Re ($NH_3ReO_4$), and S ($NH_2CSNH_2$) or adsorbed from atmosphere and subsequently diffused inside $MoS_2$. They end up incorporated within the $MoS_2$ as well with average compositions of 2.4 and 6.4 at.%, respectively. The incorporation of C in $MoS_2$ is reported to facilitate the transfer of the photo-generated electrons and holes that makes the photocatalytic reaction more efficient enhancing photocatalytic reaction[67] likewise the N incorporation improves the electronic conductivity of $MoS_2$ showing outstanding performance for hydrogen evolution reaction.[68] However, it was not possible to detect such elements locally at relatively low concentration with surface science spectroscopy. In contrast, our approach of using APT and TEM allows to obtain ppm-level chemistry information with high spatial resolution for 2D materials.

In conclusion, direct imaging of the as-synthesized Re-doped $MoS_2$ nanosheets is done using TEM and APT. The Re dopant is shown to have been incorporated within the $MoS_2$ structure, with an average composition of 3.8 at.%, and we clearly observe other impurities from heavy (V, W) to light element (C, N, O, Na), with Na showing a tendency for clustering. Our results indicate that elements from the precursor were incorporated into the nanosheets during its synthesis. The chemistry of the $MoS_2$ synthesized using wet-chemical method is much more complex than expected and often reported. Controlling the synthesis environment is crucial to avoid contamination. We expect that our approach, amenable to other nanomaterials, will help understanding the role of dopant and impurity elements on the activity of $MoS_2$ nanosheets.

**Experimental Section**

*Re-doped and non-doped $MoS_2$ nanosheet synthesis*

0.989 g of $(NH_4)_6Mo_7O_{24} \cdot 4H_2O$ (Sigma-Aldrich (Germany), ACS reagent, 99.98% trace metals basis) and 2.284 g of $NH_2CSNH_2$ (Sigma-Aldrich (Germany), ACS reagent, ≥99.0%) were dissolved in 10 ml of distilled water along with or without 0.376 g of $NH_4ReO_4$ (Sigma-Aldrich, 99.999% trace metals basis) and a homogeneous transparent solution was prepared. Then, the solution was poured into a Teflon container placed in a stainless-steel autoclave and heated at a fixed temperature of 200 °C. After 20h, the autoclave was cooled down to room temperature and the black powder was collected. Then the powder was repeatedly washed thrice with ethanol and distilled water for each centrifugation and re-dispersion steps. For $MoS_2$ materials, oxygen plasma cleaning for impurities removal of residuals could not be used since the O plasma induces substitutional oxidation of $MoS_2$ resulting MoSO/$MoO_3$ solid solution crystal.[60] Heat treatment in calcination method for removals could result MoC from as-synthesized $MoS_2$[69] and this surface cleaning method could not efficiently remove all C-based molecules.[70] Therefore, for removal of residuals, we choose the water-washing treatment.[71] Finally, the rinsed powder was dried at room temperature.

*Sample preparation: co-electrodeposition process*

As-synthesized $MoS_2$ nanosheets were electrodeposited within Ni film for embedding nanoparticles according to Kim et al.[72] Nickel sulfate hexahyrdrate ($NiSO_4$ $6H_2O$, Sigma-Aldrich (Germany)), and boric acid ($H_3BO_3$, Sigma-Aldrich (Germany)) were dissolved in distilled water. The nanosheets were then dispersed in as-prepared electrolyte solution and poured in a vertical cell for co-electrodeposition process. The vertical cell including a Cu substrate and a Pt-mesh counter electrode was used. A positive bias was applied to Pt electrode and electrodeposition of $MoS_2$ nanosheets and Ni were performed at constant current of -19 mA for 500 sec.

*TEM analysis*

TEM was performed to investigate morphology and crystal phase of Re-MoS$_2$ using 60-300 Titan Themis operated at 300 kV with a Cs-corrector for the image forming lens. Chemical composition was analyzed using EDS in STEM mode (60-300 Titan Themis operated at 300 kV with a Cs-corrector for the probe).

*XPS analysis (ask Olga)*

*APT analysis*

Needle-shaped APT specimens were prepared from MoS$_2$/Ni composite film using focused ion beam (FIB) (Helios 600) according to Thompson et al.[42] CAMECA LEAP 5000 XS system in pulsed laser mode at a specimen temperature of 50K was used for nanosheets APT analysis. A laser pulse energy of 80 pJ and a pulse frequency of 125 kHz were set. Data reconstruction was done using the Imago visualization and analysis system (IVAS) 3.8.4 developed by CAMECA instruments. The standard voltage protocol was used for all data-set reconstruction.[73]

**Supporting Information**
Supporting Information is available from the Wiley Online Library or from the author.

**Author Contributions**
[+]S.-H.K. and J.L. contributed equally.
S.-H.K. performed the co-electrodeposition, atom probe specimen preparation and atom probe analysis, with support from L.T.S. and B.G.. S.-H.K. and B.G. drafted the manuscript. J.L. performed the synthesis. J.L. and R.S performed TEM with support from C.S.. O.K. performed the XPS. All authors then contributed and have given approval to the final version of the manuscript.

**Acknowledgements**
S.-H.K., L.T.S., and B.G. acknowledge financial support from the ERC-CoG-SHINE-771602. J.L. acknowledges for the financial support from Alexander von Humboldt Foundation.

**Direct imaging of dopant and impurity distributions in 2D MoS$_2$**

*Se-Ho Kim[+], Joohyun Lim[+*], Rajib Sahu, Olga Kasian, Leigh T. Stephenson, Christina Scheu, Baptiste Gault[*]*

Se-Ho Kim, Dr. Joohyun Lim, Dr. Rajib Sahu, Dr. Olga Kasian, Dr. Leigh T. Stephenson, Prof. Christina Scheu, Dr. Baptiste Gault
Max-Planck-Institut für Eisenforschung GmbH, Max-Planck-Straße 1, Düsseldorf, 40237, Germany
E-mail: j.lim@mpie.de, b.gault@mpie.de

Dr. Baptiste Gault
Department of Materials, Royal School of Mines, Imperial College, Prince Consort Road, London, SW7 2BP, United Kingdom

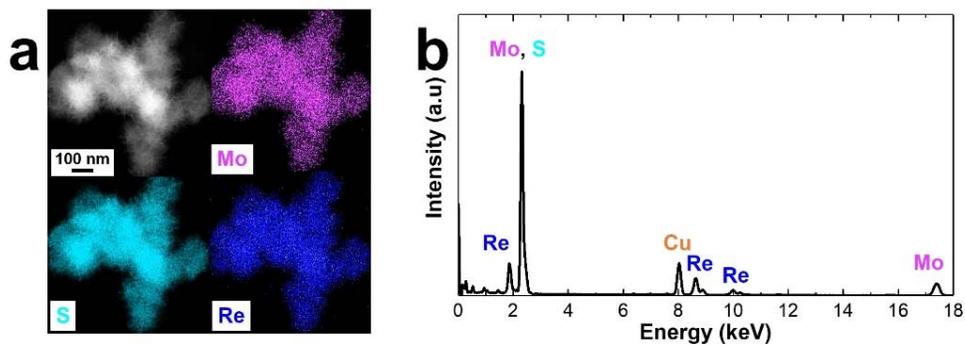

**Figure S1.** (a) (S)TEM-EDS maps and (b) EDS spectrum of Re-doped MoS$_2$ on a Cu TEM grid.

The overlapped peak at 2.30 keV from the acquired EDS spectrum corresponds to both Mo (2.29 keV) and S (2.31 keV) signals, while the peak at 1.85 keV indicates that the Re (1.84 keV) atoms are presented within the Re-MoS$_2$. Signals below 1.5 keV are too weak that the corresponding elements are difficult to confirm.

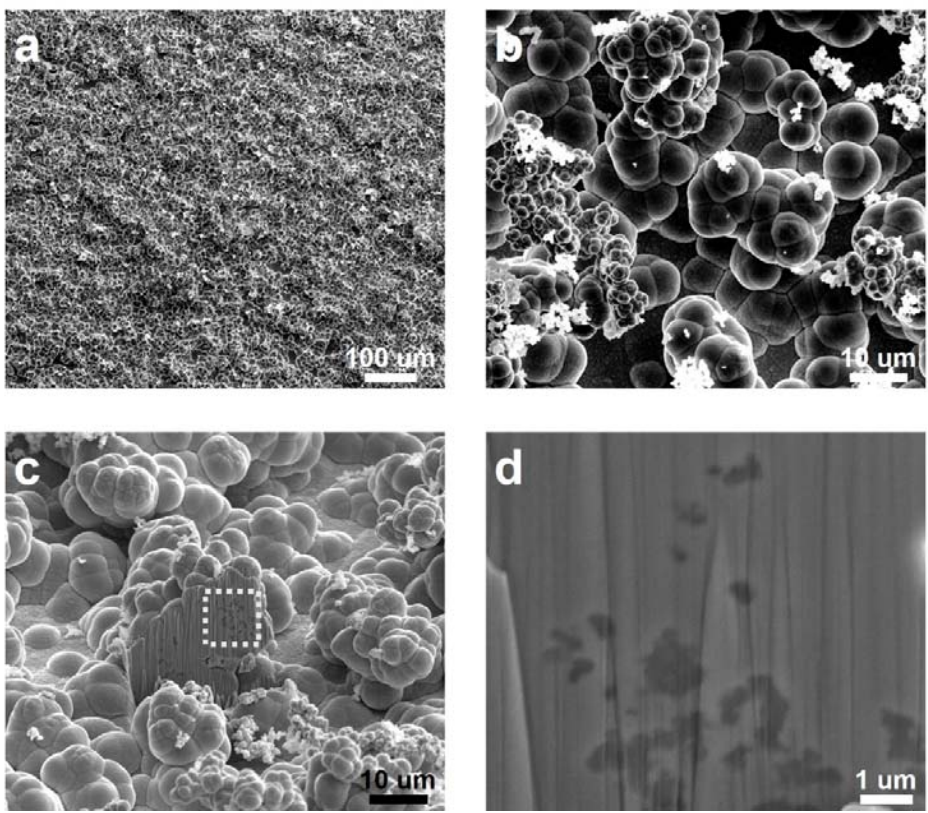

**Figure S2.** (a) and (b) Surface FIB-scanning electron microscopy (SEM) images of co-electrodeposited Ni/Re-MoS$_2$ sample. (c) Cross-sectional FIB-SEM image tilted 52° of co-deposited Ni/Re-MoS$_2$ sample. (d) High resolution image of the sectioned area from Figure 2c (dotted square box).

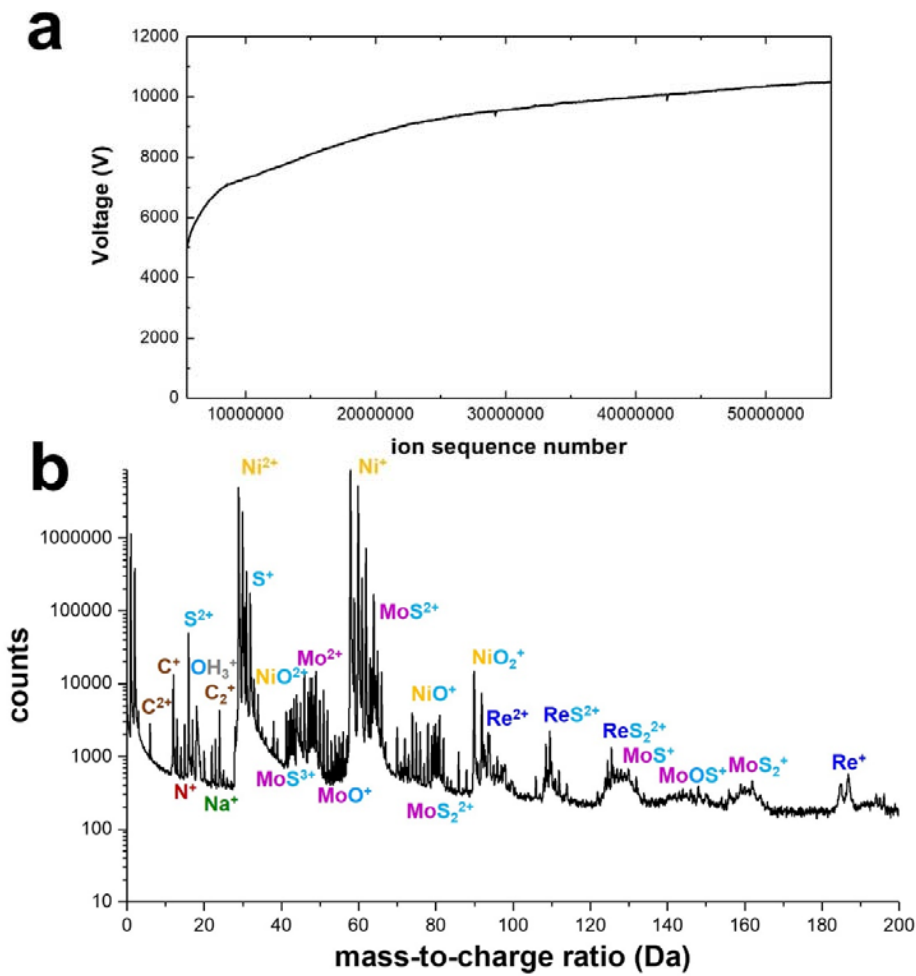

**Figure S3.** (a) Voltage *vs.* ion sequence number curve during the APT measurement. (b) Acquired overall mass spectrum data.

During the APT measurement, there was no sign of a strong voltage- drop or increase indicating that a field ion evaporation behavior was stable for the Re-MoS$_2$ embedded in Ni (see Figure S3a). In the overall mass spectrum, major isotope peaks can be assigned to MoS$_x$ and Ni in single and double charge states as shown in Figure S3b. The intentional doping element, Re, is detected at 185 and 187 Da for single charge state (Re$^+$) and 92.5 and 93.5 Da for double charge state (Re$^{2+}$) in the mass spectrum. Also, ReS$_x^+$ molecular ions are detected which could support that Re atoms are substitutional dopant in MoS$_2$ rather than interstitial element as reported from the density functional theory (DFT) calculation and (S)TEM result.[1,2]

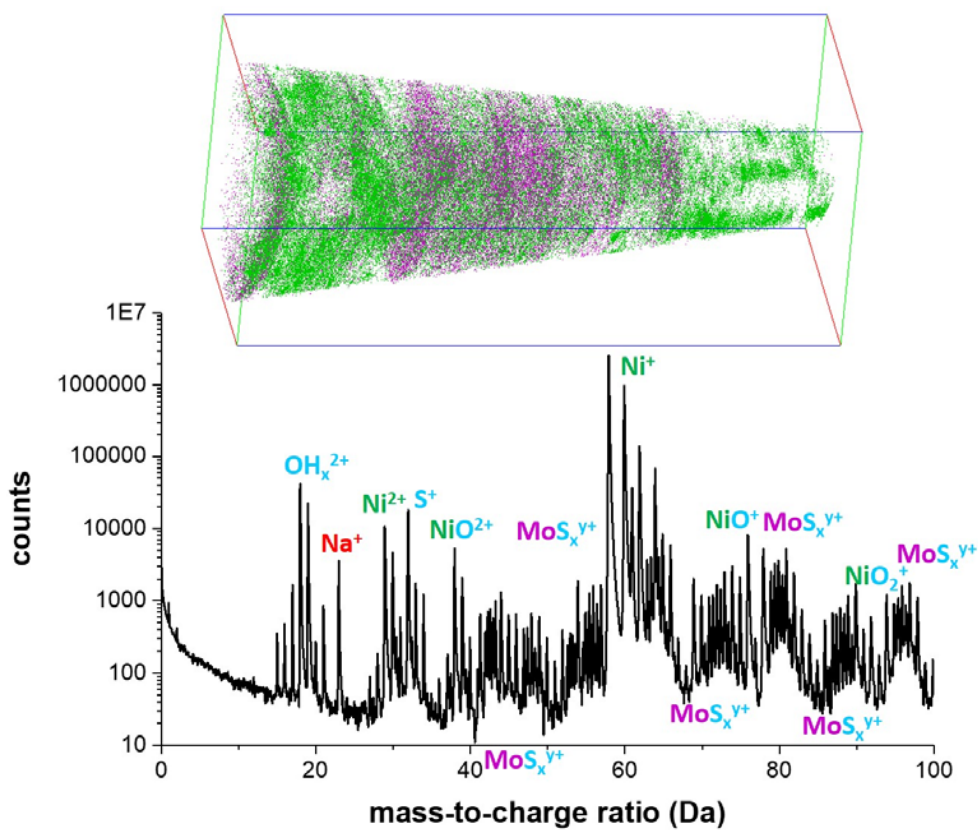

**Figure S4.** 3D reconstruction of non-doped $MoS_2$ nanosheets in Ni with the corresponding mass spectrum. (up) Reconstructed 3D atom map. Green and purple dots represent the reconstructed Ni and Mo atoms, respectively. (down) The corresponding mass spectrum.

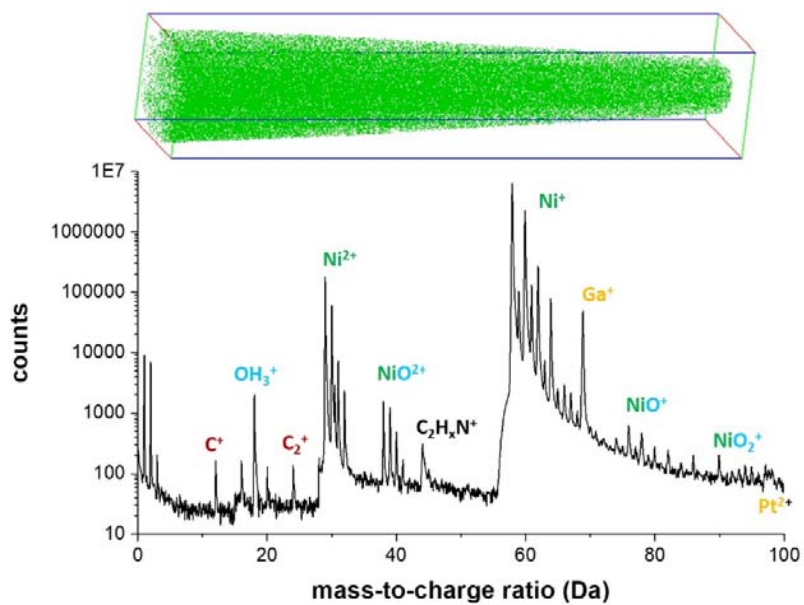

**Figure S5.** APT results of Re-doped MoS$_2$ in Ni where no Re-doped MoS$_2$ detected. (up) Reconstructed 3D atom map. Green dots represent the reconstructed Ni atoms. (down) The corresponding mass spectrum.

**Table S1.** List of the reconstructed Na clusters with surface area (nm$^2$) and diameter (nm).

| No. cluster | Iso-surface Na at 0.049 at.% | |
|---|---|---|
| | Surface Area (nm$^2$) | Diameter (nm) |
| 1 | 9.16 | 1.71 |
| 2 | 25.81 | 2.87 |
| 3 | 10.73 | 1.85 |
| 4 | 33.38 | 3.26 |
| 5 | 59.23 | 4.34 |
| 6 | 109.55 | 5.91 |
| 7 | 130.12 | 6.44 |
| 8 | 19.63 | 2.50 |
| 9 | 290.15 | 9.61 |
| 10 | 49.32 | 3.96 |
| 11 | 43.68 | 3.73 |
| 12 | 22.57 | 2.68 |
| 13 | 38.84 | 3.52 |
| 14 | 17.97 | 2.39 |
| 15 | 107.94 | 5.86 |
| 16 | 12.96 | 2.03 |
| 17 | 122.57 | 6.25 |
| 18 | 128.57 | 6.40 |
| 19 | 17.21 | 2.34 |
| 20 | 20.54 | 2.56 |
| 21 | 63.35 | 4.49 |
| 22 | 502.41 | 12.65 |
| 23 | 26.62 | 2.91 |
| 24 | 65.37 | 4.56 |
| 25 | 482.95 | 12.40 |
| 26 | 29.67 | 3.07 |
| 27 | 3.53 | 1.06 |
| 28 | 231.17 | 8.58 |
| 29 | 9.52 | 1.74 |
| 30 | 165.68 | 7.26 |
| 31 | 90.88 | 5.38 |
| 32 | 19.46 | 2.49 |
| 33 | 11.19 | 1.89 |
| 34 | 102.84 | 5.72 |
| Average diameter (nm) | | 4.5 |

From the IVAS software, the cluster analysis of Na was performed from Re-doped MoS$_2$ nanosheet data. Iso-surface of Na at 0.049 at.% was selected as a threshold value for Na (bulk atomic concentration of 0.097 at.%) to highlight the interface between the $MoS_2$ and Na clusters. Spherical shape assumption was made for Na clusters.

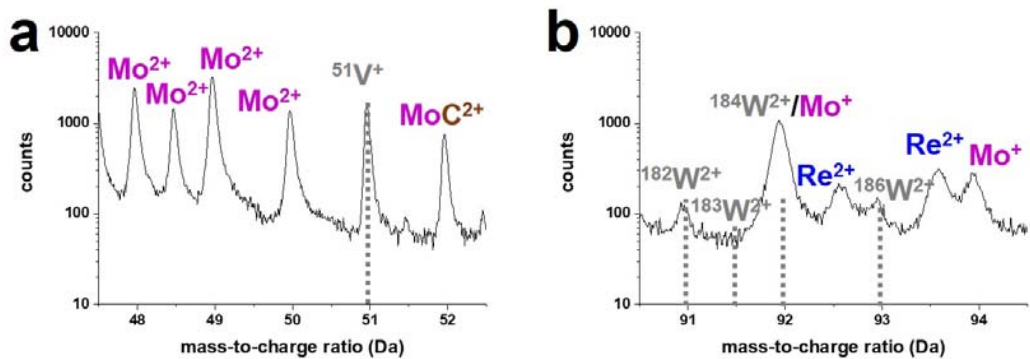

**Figure S6.** Sectioned mass spectra of Re-MoS$_2$ result from Figure S3b for (a) a V$^+$ peak and (b) W$^{2+}$ peaks.

**Table S2.** Isotope atomic percentage for natural isotopes and measured isotopes of W acquired from the peak decomposition bulk analysis.

| Stable W (at.%) | W isotope | Detected W (at.%) |
| --- | --- | --- |
| 27.0 | $^{182}$W | 23.5 |
| 14.3 | $^{183}$W | 22.0 |
| 30.6 | $^{184}$W | 32.5 |
| 28.4 | $^{186}$W | 21.9 |

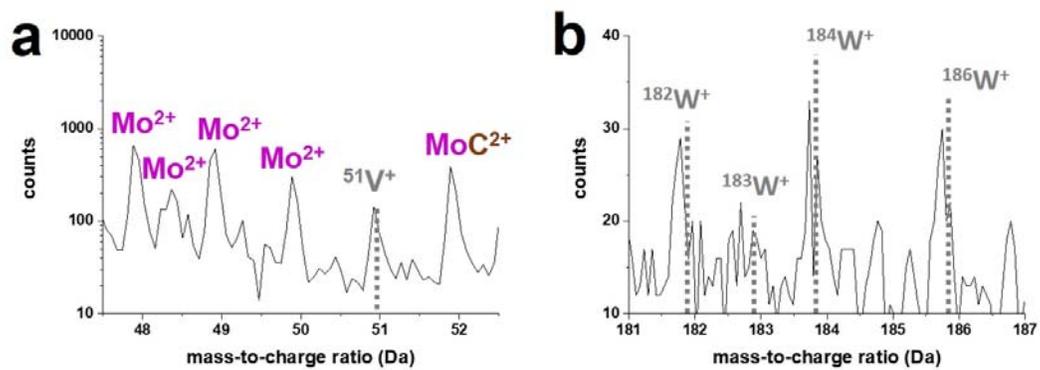

**Figure S7.** Sectioned mass spectra of non-doped $MoS_2$ result from Figure S4b for (a) $^{51}V^+$ peak and (b) $W^+$ peaks.

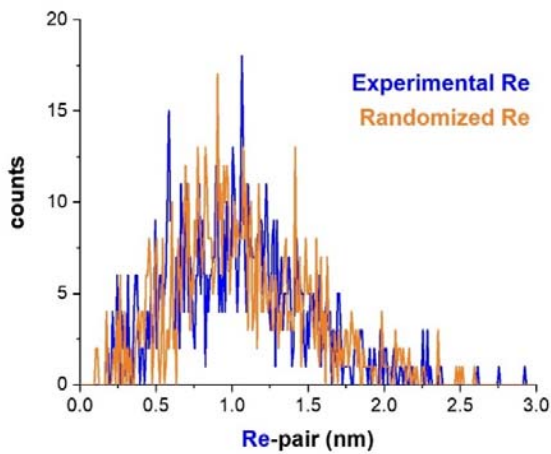

**Figure S8.** Re-Re nearest-neighbor distance distribution in Re-doped MoS$_2$ nanosheets (sample width Re-pair = 0.01 nm)

In order to quantify whether Re exhibits a tendency to segregate, we performed a Re-Re nearest-neighbor analysis[3] on an individual nanosheet that was first extracted from within a Re-doped MoS$_2$ nanosheets dataset. The distribution was compared with a randomized Re distribution, in which the atomic positions are unchanged by the mass-to-charge ratios are randomly swapped. Figure S8 shows that the experimental Re-Re nearest-neighbor distribution has no significant deviation from the randomized Re distribution. A frequency distribution analysis was also performed, and the corresponding Pearson coefficient ($\mu$), which gives a quantitative assessment of the randomness associated to a a $\chi^2$-statistical test.[4] The coefficient value lies between 0 and 1, where $\mu = 0$ corresponds to complete randomness and $\mu = 1$ is completely ordered. The calculated $\mu_{Re}$ value acquired from the Re-MoS$_2$ nanosheets data is 0.044, which qualitatively proves that Re is homogenously distributed within the as-synthesized Re-MoS$_2$. These analyses were performed within the IVAS 3.8.4 software suite.

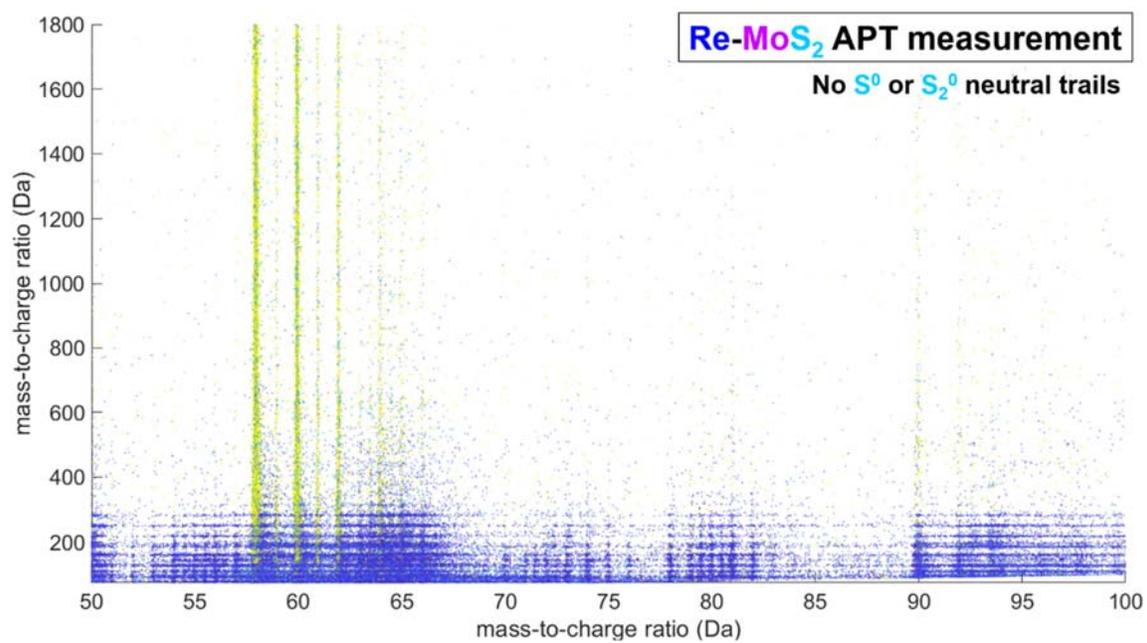

**Figure S9.** Correlation histograms for Re-MoS$_2$/Ni field evaporation. Note that there is no significant neutral and molecular-ion dissociation trails.

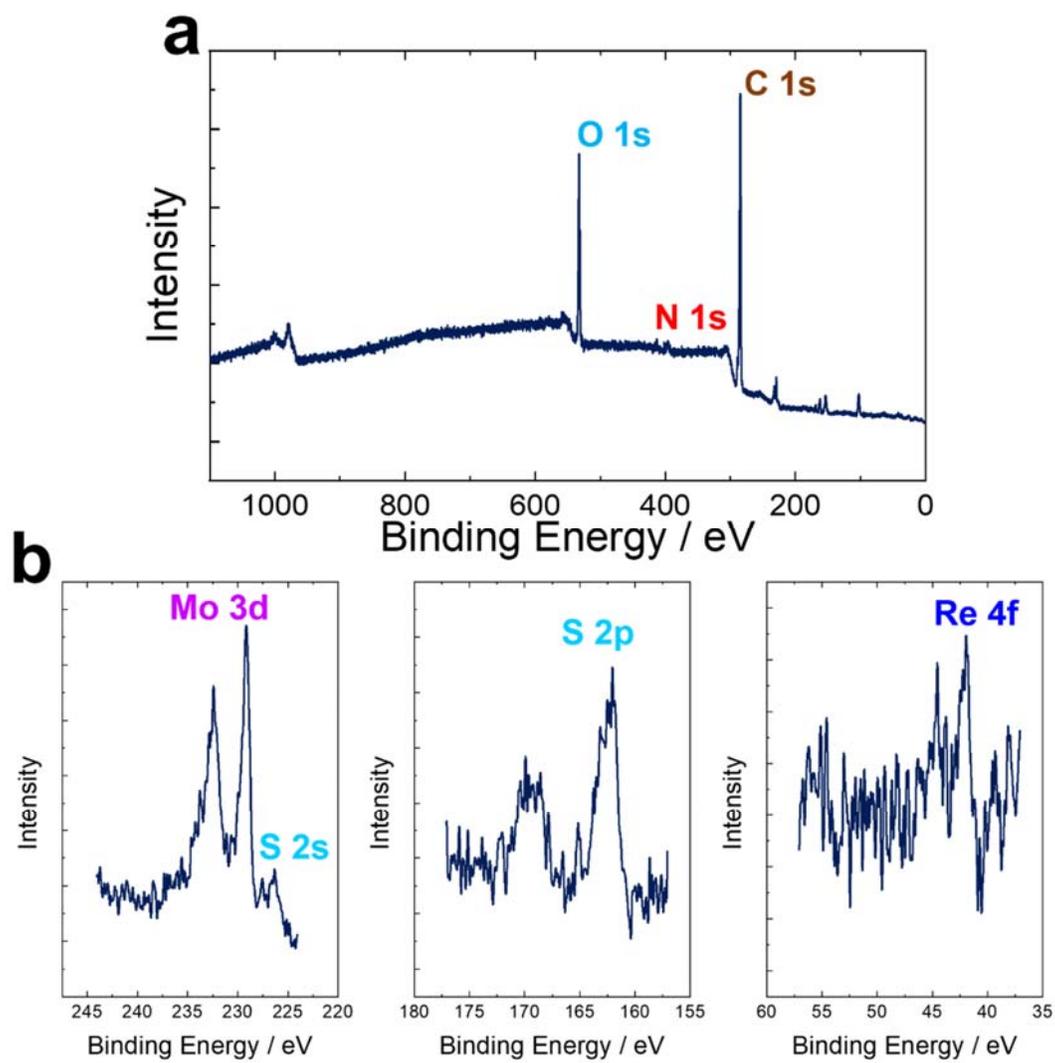

**Figure S10.** (a) XPS survey spectrum of Re-MoS$_2$ sample. (b) Sectioned spectra in three different ranges for Mo, S, and Re.

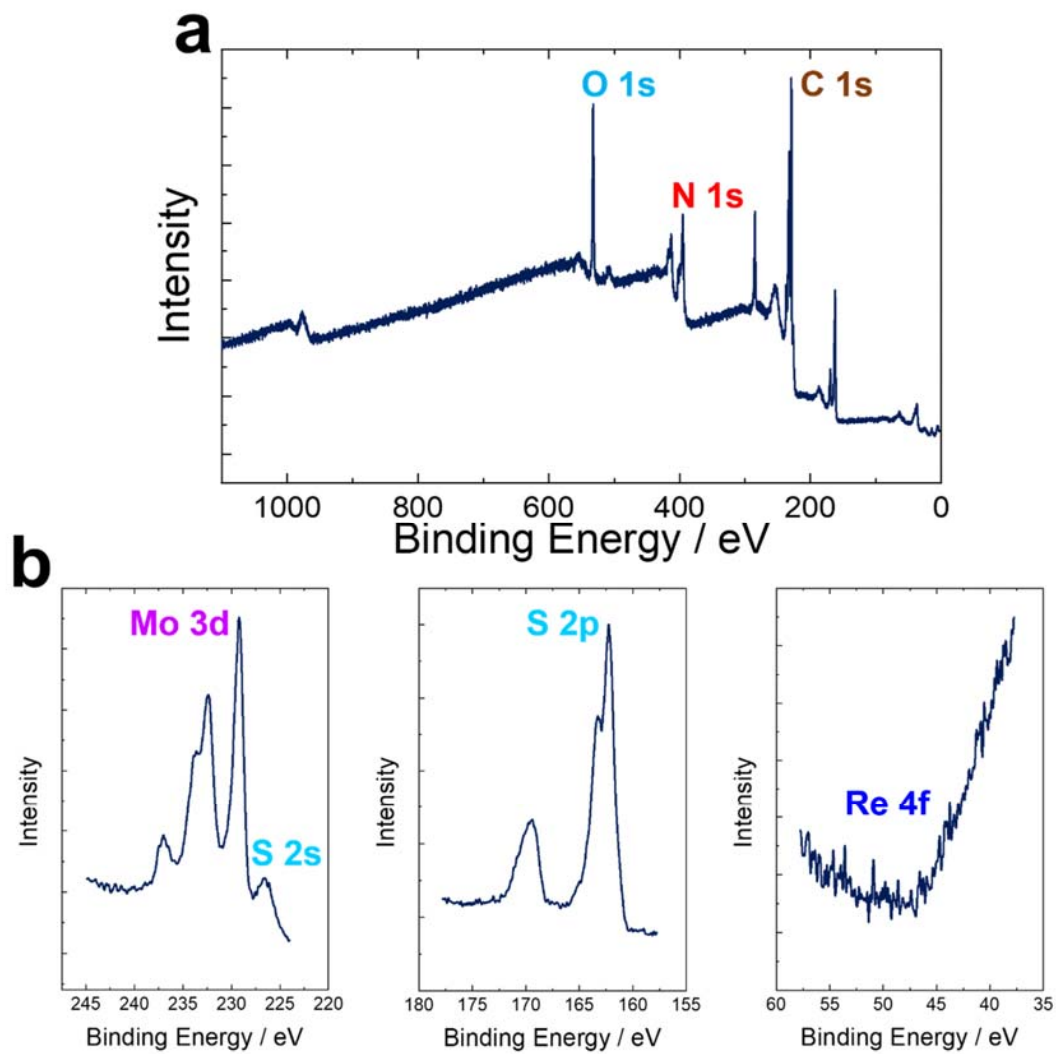

**Figure S11.** (a) XPS survey spectrum of non-doped MoS$_2$ sample. (b) Three different core level spectra of Mo, S, and Re (see Figure S12 for the Mo 3d peak deconvolution).

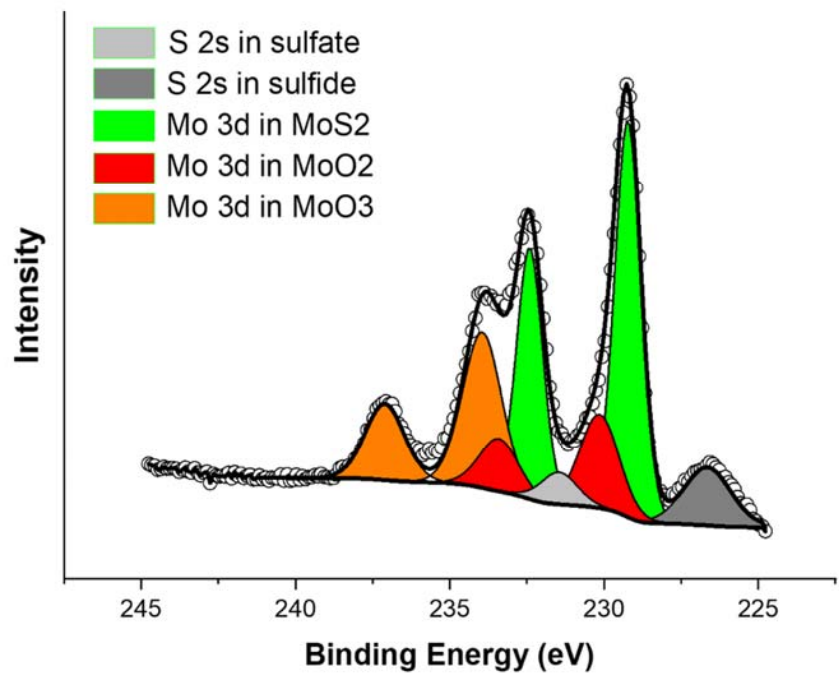

**Figure S12.** MoS$_2$ nanosheet analysis of XPS. Deconvolution of the Mo 3d core level spectra.